\documentclass[twocolumn,prl,secnumarabic,amssymb,amsmath,nobibnotes,aps,showpacs]{revtex4-1}

\usepackage{graphics} \usepackage{graphicx} \usepackage{braket} \usepackage{float} \usepackage{tabularx} % Advanced table configurations
\usepackage[hyperindex,breaklinks]{hyperref}
\usepackage[usenames,dvipsnames]{color}

\begin{document}

\title{In-vacuum scattered light reduction with cupric oxide surfaces for sensitive fluorescence detection}

%\author{E. B. Norrgard$^{1}$, N. Sitaraman$^{1}$, and D. DeMille$^{1}$}

\author{E. B. Norrgard}

\email[Electronic address: ]{eric.norrgard@yale.edu}
\author{N. Sitaraman}
\altaffiliation[Current address: ]{Department of Physics, Cornell University, 142 Sciences Dr., Ithaca, New York 14853, USA}

\author{J. F. Barry}
\altaffiliation[Current address: ]{Harvard-Smithsonian Center for Astrophysics, 60 Garden Street, Cambridge, Massachusetts 02138, USA.}

\author{D. J. McCarron}

\author{M. H. Steinecker}

\author{D. DeMille}

%\altaffiliation[Current address: ]{Department of Physics, Cornell University, 142 Sciences Dr., Ithaca, New York 14853, USA}

\affiliation{Department of Physics, Yale University, P.O. Box 208120, New Haven, Connecticut 06520, USA}

\begin{abstract} We demonstrate a simple and easy method for producing low-reflectivity surfaces that are ultra-high vacuum compatible, may be baked to high temperatures, and are easily applied even on complex surface geometries.  Black cupric oxide (CuO) surfaces are chemically grown in minutes on any copper surface, allowing for low-cost, rapid prototyping and production.  The reflective properties are measured to be comparable to commercially available products for creating optically black surfaces.  We describe a vacuum apparatus which uses multiple blackened copper surfaces for sensitive, low-background detection of molecules using laser-induced fluorescence. \end{abstract}


\maketitle

 %Introduction
 \section{Introdution}
Laser-induced fluorescence (LIF) detection is a simple yet powerful technique frequently used in gas-phase atomic and molecular spectroscopy \cite{Kinsey1977,Scoles1988,Pauly2000,Demtroder2015}.  In many interesting situations, LIF signals are fairly small. For example, in the gas phase, rovibrational branching from molecular electronic states may lead to LIF signals of $\lesssim$\,1 photon per molecule.  Modern experiments with atoms \cite{Gross2015}, molecules \cite{Moerner2003}, and ions \cite{Leibfried2003} frequently aim to detect small numbers of particles---sometimes as few as one.  In such cases, scattered laser light (and the noise from it) can seriously degrade the signal-to-background and signal-to-noise ratios in LIF detection.

   Blackened surfaces are often used to suppress scattered light in LIF detection experiments \cite{Scoles1988, Hall1984, Pauly2000}.  Recently, we have investigated optically black materials for stray light suppression to facilitate efficient detection of molecules via LIF  (for use in experiments on laser slowing \cite{Barry2012}, cooling \cite{Shuman2010}, and trapping \cite{Barry2014, McCarron2015,Norrgard2015} of SrF molecules).  Here, we present a method for producing ultra-high vacuum (UHV) compatible low-reflectivity surfaces that is simple to implement and very flexible in its utility.  Our method, which we refer to as ``copper blackening'', uses a simple chemical treatment to produce black, dendritic cupric oxide (CuO) on the surfaces of any pre-formed copper part.  We describe the method and present measurements of the reflectance properties of these prepared surfaces.  Measurements on blackened copper surfaces are compared to measurements on well-characterized commercial surfaces.

   Blackened copper has a number of attractive properties for stray light suppression solutions.  Copper is easily machined or formed, and the surfaces may be grown in minutes on parts of any manufactured geometry.  Vacuum components may be made of blackened copper, and undesired reflective surfaces may be covered with blackened copper foils or sheets formed into any desired shape.   Blackened copper is suitable in UHV from cryogenic temperatures up to 500\,$^{\circ}$C \cite{Kirsch2001}.

\section{Blackening Method}

 Copper has two common oxidization states: the red $\rm{Cu_2O}$ (cuprous oxide) and the black CuO (cupric oxide).  The black CuO is effectively grown on the surface of metallic copper by immersing the metal in a solution composed of 100 g of NaOH and 100 g of NaCl$\rm{O_2}$ per liter of deionized water \cite{Meyer1944}.  The final step of the reaction,
 \begin{equation} \rm{Cu(OH)_2\,(s) \xrightarrow[heat]{} CuO\,(s) + H_2O\,(l),} \label{eq:reaction} \end{equation}
 grows the black CuO \cite{Filipic2012}.  The solution's temperature is maintained between 95\,$^\circ$C and 100\,$^\circ$C during the blackening process.  Thermal and concentration gradients are minimized by using a magnetic stir bar in the blackening solution.  Immersing the beaker holding the blackening solution in a heated outer water bath also reduces thermal gradients.  The outer bath is saturated with NaCl to allow its temperature to exceed 100\,$^\circ$C without boiling and potentially contaminating the solution.

The material used for all parts described in this paper is oxygen-free high conductivity (OFHC) copper (alloy 101), selected for its desirable UHV properties.  As described below, we tested OFHC copper prepared with surfaces both as-finished, and sandblasted to increase the roughness.  We have also performed the blackening process on alloy 110 Cu and on brass parts, which might be useful in applications where UHV is not a concern (while we have not measured the reflectance properties of these parts, they appear similarly black by eye to the 101 alloy measured in this work).  Most steels and titanium are minimally corroded by the solution \cite{Sandvik} and may be used to hold or remove parts from the solution, but aluminum rapidly dissolves in NaOH.

Prior to blackening, parts are cleaned in an ultrasonic bath of $\sim$\,1\,\% Citranox in deionized water for one hour to remove any surface contamination, followed by a rinse in a deionized water bath.  Parts are then immediately placed in the blackening solution for 10 minutes (by eye, parts blackened for 10 minutes are not distinguishable from parts blackened for 5 or 15 minutes; under similar conditions, the oxide layer thickness was observed to saturate in 2.5 minutes in Ref.\,\cite{Lebbai2003}).  Parts are then removed from the solution, rinsed in a fresh deionized water bath, and placed in an ultrasonic acetone bath for one hour.  After a final fresh deionized water rinse, parts are dried with a gentle stream of dry N$_2$.

CuO is known to grow on surfaces in a microstructure of fine dendrites by a number of wet-chemical \cite{Filipic2012} and electrochemical methods \cite{Shao2009, Hu2010}.  These dendrites effectively trap incident light \cite{Xia2014}, further reducing reflection.  Care must be taken when handling the blackened parts, as the CuO dendrites are easily crushed by the pressure of touching the surface, rinsing with a squirt bottle, or drying with a high pressure of compressed gas.  Crushing the dendrites is observed to result in a locally less absorptive surface.  However, the dendrites are not observed to flake from the surface when crushed, or when subjected to mechanical or ultrasonic vibration.

In some instances, having a part with only some areas blackened is desirable.  While we find the thin CuO layer may be easily removed by mechanical means, such as sanding or machining, it is not always possible to perform such operations without crushing the dendrites in the areas where blackening is desired.  We have investigated techniques for masking areas to inhibit CuO growth, and found adhesive polytetrafluoroethylene (PTFE) tape to provide good results (see below).  PTFE does not react with the blackening solution, and the acrylic adhesive leaves little visible residue; what does remain is removed in the ultrasonic acetone bath typically used in our cleaning procedure.  The blackening solution creeps only a few mm under the edges of the tape during a 10 minute blackening, leaving a reasonably sharp boundary.

\section{Method for Measuring Reflectance}

\begin{figure}[b]
\centering
\includegraphics[width= \linewidth]{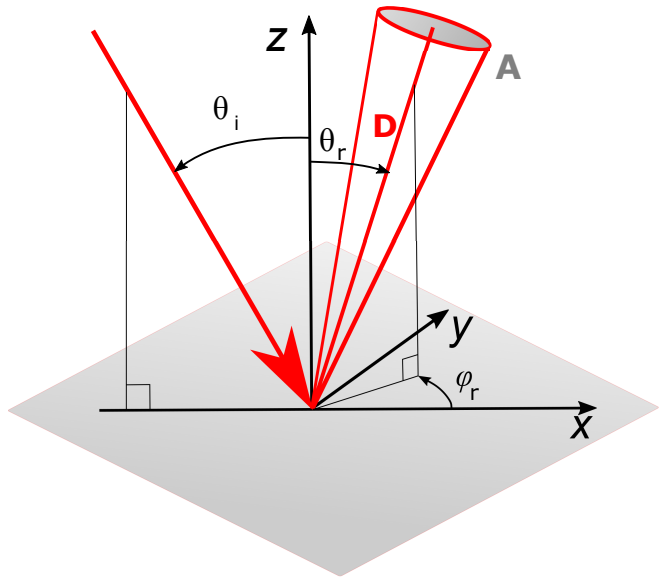}
        \caption{(Color online) Coordinate system used in measurements.  The $z$-axis is the surface normal.  Incident light travels in the $x-z$ plane at angle $\theta_{\rm{i}}$ to the normal.  Reflected light is measured at polar angle $\theta_{\rm{r}}$ and azimuthal angle $\phi_{\rm{r}}$ on a detector of area $A$ a distance $D$ away.}\label{fig:geometry}
\end{figure}

  When attempting to eliminate background light, the relative positions of the light source, scattering surface, and detector can significantly affect the relative intensity of scattered light.  For example, surfaces typically have dramatically different reflectance properties near normal incidence and near grazing incidence.  Because of this, optical reflectance is typically described in terms of the bidirectional reflectance distribution function (BRDF) $f_{\rm{r}}(\theta_{\rm{i}}, \theta_{\rm{r}}, \phi_{\rm{r}})$, defined as
\begin{equation}
f_{\rm{r}}(\theta_{\rm{i}}, \theta_{\rm{r}}, \phi_{\rm{r}}) = \frac{P_{\rm{r}} D^2}{P_{\rm{i}} A \cos \theta_{\rm{r}}}.
\label{brdf}
\end{equation}
  Here, $P_{\rm{r}}$ is the power reflected to the detector from incident power $P_{\rm{i}}$.  The geometrical quantities used to define $f_{\rm{r}}$ are depicted in Fig.\,\ref{fig:geometry}: $A$ is the area of the detector, $D$ is the distance from the scattering surface to the detector, $\theta_{\rm{i}}$ and $\theta_{\rm{r}}$ are the incident and reflected polar angles, respectively, and $\phi_{\rm{r}}$ is the relative reflected azimuthal angle.

If the geometric collection efficiency for the scattering source is large (i.e. the detector subtends a large solid angle), or if there are multiple scattering surfaces at different angles relative to the incident light, the directional-hemispherical reflectance (DHR) $\rho(\theta_{\rm{i}})$, defined as
 \begin{equation}
\rho(\theta_{\rm{i}}) = \int f_{\rm{r}}(\theta_{\rm{i}}, \theta_{\rm{r}}, \phi_{\rm{r}})\cos(\theta_{\rm{r}}) d\Omega_{\rm{r}},
 \label{dhr}
  \end{equation}
  is a more useful characteristic of the surface.

For a simple relative comparison of materials, we direct a laser beam with 1/$e^2$ diameter $\approx$\,1\,mm to intersect the axis of rotation of a turntable, with the sample of interest at its center.  The sample and detector photodiode may be rotated independently about this axis.  A chopper wheel modulates the incident light power $P_{\rm{i}}$ at 3\,kHz, and the reflected power $P_{\rm{r}}$ is extracted from the detector photodiode signal using a lock-in amplifier.
 A fraction of the beam $P_{\rm{m}}$ is sent to a monitor photodiode to account for source intensity fluctuations.   Prior to a reflectance measurement, the signal and monitor photodiodes record $P_{\rm{i}}$ and $P_{\rm{m}}$, respectively.  During the reflectance measurement, $P_{\rm{i}}$ is normalized to the monitor photodiode signal.

 The simplicity of this setup limits us to measurements with $\phi_{\rm{r}}$\,=\,0$^{\circ}$ (forward scattering) or $\phi_{\rm{r}}$\,=\,180$^{\circ}$ (back scattering).  To describe these data, we define the bidirectional azimuth-independent reflection distribution \linebreak

 \begin{figure*}[t]
\centering
\includegraphics[width= .835\textwidth]{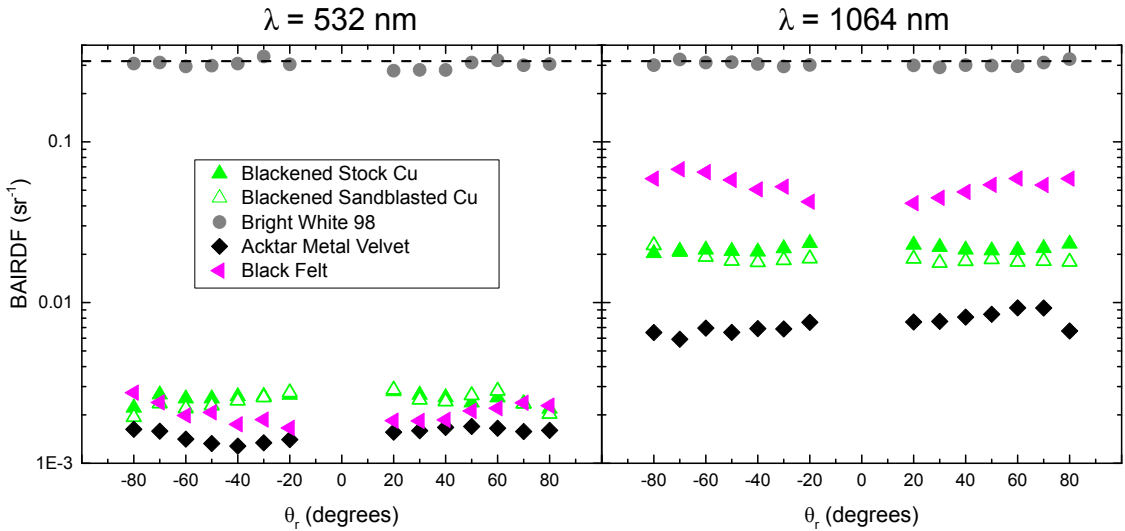}
        \caption{(Color online) Bidirectional azimuth-independent reflection distribution function vs reflection angle for materials investigated with $\lambda$\,=\,532\,nm (left) and $\lambda$\,=\,1064\,nm (right) at normal incidence ($\theta_{\rm{i}}$\,=\,0$^\circ$).  The dashed line marks a Lambertian surface with unit reflectivity.}\label{fig:material}
\end{figure*}

\begin{figure*}[t]
\centering
\includegraphics[width= .835\textwidth]{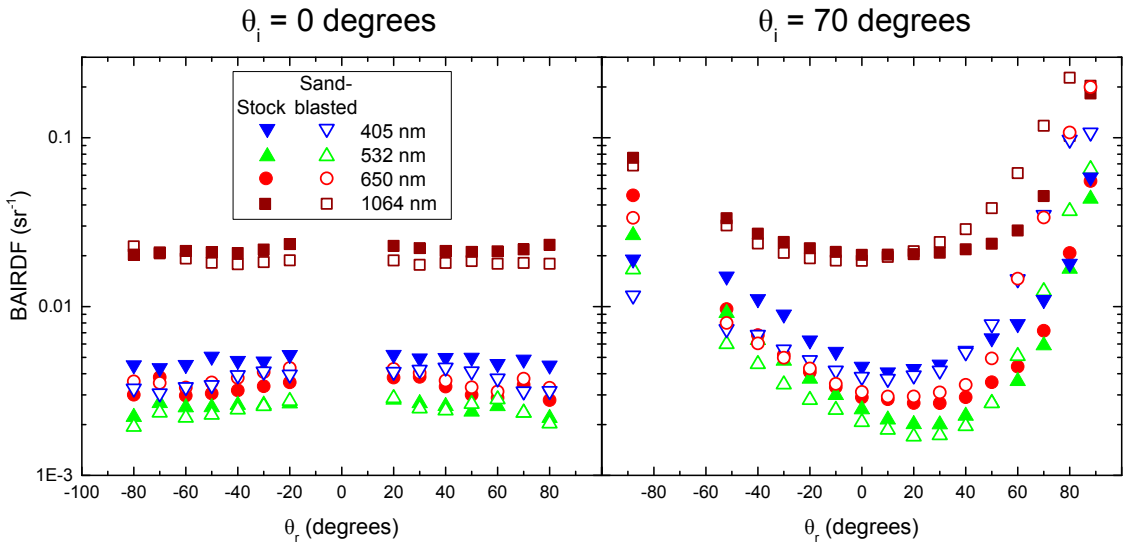}
        \caption{(Color online) Bidirectional  azimuth-independent reflection distribution function vs reflection angle for blackened stock copper (filled markers) and blackened sandblasted copper (open markers) at normal incidence (left) and 70$^{\circ}$ incidence (right).}\label{fig:angle}
\end{figure*}

\begin{table}[h]

 \begin{center} \begin{tabularx}{\columnwidth}{l X c X c} \toprule

Material& & $\lambda\,=\,$532\,nm & &  $\lambda\,=\,$1064\,nm \\ \hline
Bright White 98 & & 0.95 & &0.95\\
Blackened Cu, Stock & &0.008 & &0.07 \\
Blackened Cu, Sandblasted & & 0.008& &0.06\\
 Black Felt & & 0.006& &0.17\\
 Acktar Metal Velvet & & 0.005& &0.024\\ \toprule

\end{tabularx}\par
\caption{Directional-hemispherical reflection $\rho(\theta_{\rm{i}}=0^{\circ})$ for materials investigated.}
\label{tab:material}
\end{center}
\end{table}

\begin{table}[h]

 \begin{center} \begin{tabularx}{.75 \columnwidth}{l X c X c} \toprule

 $\lambda$ & &Stock & & SandBlasted\\\hline
 405\,nm & & 0.015& &0.012\\
  532\,nm & &0.008& &0.008  \\
  650\,nm & &0.010& &0.013\\
   1064\,nm & & 0.07& &0.06\\ \toprule

\end{tabularx}\par
\caption{Directional-hemispherical reflection $\rho(\theta_{\rm{i}}=0^\circ)$ for blackened stock copper and blackened sandblasted copper vs wavelength $\lambda$.}
 \label{tab:angle}
 \end{center}
 \end{table}

 \hfill
 \linebreak
 \hfill
 \linebreak
  function (BAIRDF) as $F_{\rm{r}}(\theta_{\rm{i}}, \theta_{\rm{r}})$\,=\,$f_{\rm{r}}(\theta_{\rm{i}}, \theta_{\rm{r}}, \phi_{\rm{r}})$,  with $\theta_{\rm{r}}$ positive (negative) for $\phi_{\rm{r}}$\,=\,$0^{\circ}$ ($\phi_{\rm{r}}$\,=\,$180^{\circ}$).
 To assign a value for the DHR from the BAIRDF at normal incidence ($\theta_{\rm{i}}$\,=\,0$^\circ$), we perform the integral over $\phi_{\rm{r}}$ in Eq.\,\ref{dhr} by averaging over the measured values for $\phi_{\rm{r}}$\,=\,$0^{\circ}$ and $\phi_{\rm{r}}$\,=\,$180^{\circ}$ (This is essentially equivalent to assuming azimuthally symmetric scattering for the case of normal incidence).
 
\section{Discussion of Results}
%The BRDF was measured for four common laser wavelengths $\lambda$: 405, 532, 650, and 1064 nm.
In Fig.\,\ref{fig:material}, we compare $F_{\rm{r}}$ at normal incidence for five materials at wavelengths $\lambda$\,=\,532\,nm and 1064\,nm.  Values for $\rho(\theta_{\rm{i}}$\,=\,0$^\circ)$ are tabulated in Table \ref{tab:material}.  To confirm that the calculated $F_{\rm{r}}$ and $\rho$ values were reliable, $\rho$ was measured for a near-Lambertian, high-reflectivity surface (Bright White 98).  We measured \mbox{$\rho(\theta_{\rm{i}}$\,=\,0$^{\circ})$\,=\,0.95} at 532\,nm, compared to the manufacturer-specified value 0.96 typical at 550\,nm (no test data at 1064\,nm were available from the manufacturer).  For a commercial light-absorbing blackened surface provided on aluminum foil (Acktar Metal Velvet, AMV), we measure \mbox{$\rho (\theta_{\rm{i}}$\,=\,0$^\circ$)\,=\,0.005} at 532\,nm, also in fair agreement with the manufacturer's specified typical value of 0.006.

In all cases investigated at normal incidence, the AMV provided the blackest surface.   However, the blackened copper had only marginally higher reflectance than the AMV, and proved generally comparable to standard black felt cloth.   At normal incidence, no significant difference was measured between blackened copper starting with a stock finish and copper with its surface roughened by sandblasted prior to cleaning and blackening.  At $\lambda$\,=\,532\,nm, blackened copper is $\approx$\,50\% more reflective than black felt, and  $\approx$\,75\% more reflective than AMV.  At 1064\,nm, blackened copper is $\approx$\,2$\times$ less reflective than black felt, and $\approx$\,3$\times$ more reflective than AMV.

  In Fig.\,\ref{fig:angle}, we plot the measured BAIRDF as a function of reflected angle $\theta_{\rm{r}}$ for incident angle $\theta_{\rm{i}}$\,=\,0$^{\circ}$ (left) and $\theta_{\rm{i}}$\,=\,70$^{\circ}$ (right) for four common laser wavelengths $\lambda$: 405, 532, 650, and 1064 nm.  The DHR for normal incidence is given in Table \ref{tab:angle}.  As a consistency check of our BAIRDF data, we note that by reciprocity we expect $F_{\rm{r}}(\theta_{\rm{i}}, \theta_{\rm{r}})\,=\,F_{\rm{r}}(\theta_{\rm{r}}, \theta_{\rm{i}})$.  For all cases in Fig.\,\ref{fig:angle}, these values agree to within 20\,\%.

  At normal incidence, the blackened stock and sandblasted copper display reflectance nearly independent of $\theta_{\rm{r}}$.  Both surface preparation methods produce comparable values of $\rho(\theta_{\rm{i}}\,=\,0^{\circ})$ for all wavelengths tested.

Blackened copper produced from stock material was found to specularly reflect for $\theta_{\rm{i}}\gtrsim 75^{\circ}$, making it highly ineffective at reducing scattered light at near-grazing incidence.  Sandblasting (using sand with typical particle diameter $\sim$\,10\,$\mu$m) produces a diffuse surface over a much larger scale than the typical scale of the dendrites ($\sim$\,100\,nm \cite{Xia2014,Filipic2012}).  We expected that sandblasting would tend to randomize the surface normal, and indeed no specular reflection component was observed for blackened sandblasted copper for any angle of incidence we were able to test, $\theta_{\rm{i}} < 89^\circ$.  However, the removed specular reflection is accompanied by increased forward (positive $\theta_{\rm{r}}$) diffuse scatter.  For $\theta_{\rm{i}}\,=\,70^{\circ}$ (an incident angle just below the onset of significant specular reflection in blackened stock copper), the stock finish has less forward scatter than the sandblasted copper, while back scatter is lower for the sandblasted copper.

\begin{figure}[b]
\centering
\includegraphics[width= \linewidth]{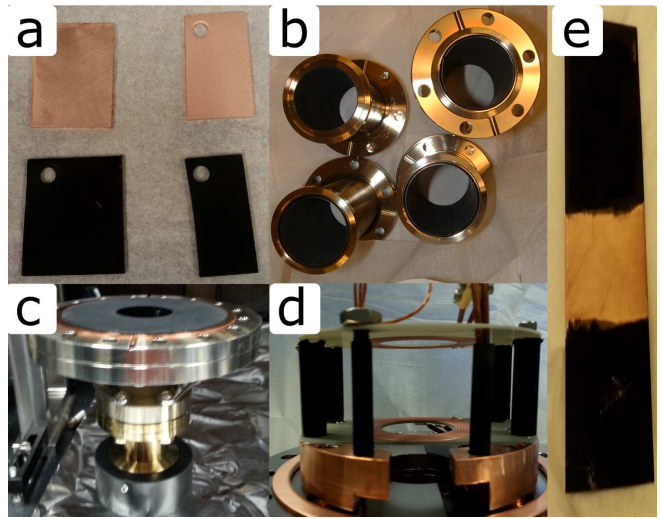}
        \caption{(Color online) Examples of blackened copper parts. (a) Stock (left) and sandblasted (right) copper samples before (top) and after (bottom) the blackening process.  (b) CF full nipples with blackened copper inserts.  (c) Blackened copper disk covers CF flange with hole in the center for optical access.  (d) Blackened copper rods serves as heat (thick outer rods) and electronics conduits (thin inner rods) for in-vacuum magnet circuit boards. (e) Copper sheet with midsection masked from the blackening process using adhesive PTFE tape.}\label{fig:examples}
\end{figure}

\begin{figure}[t]
\centering
\includegraphics[width= \linewidth]{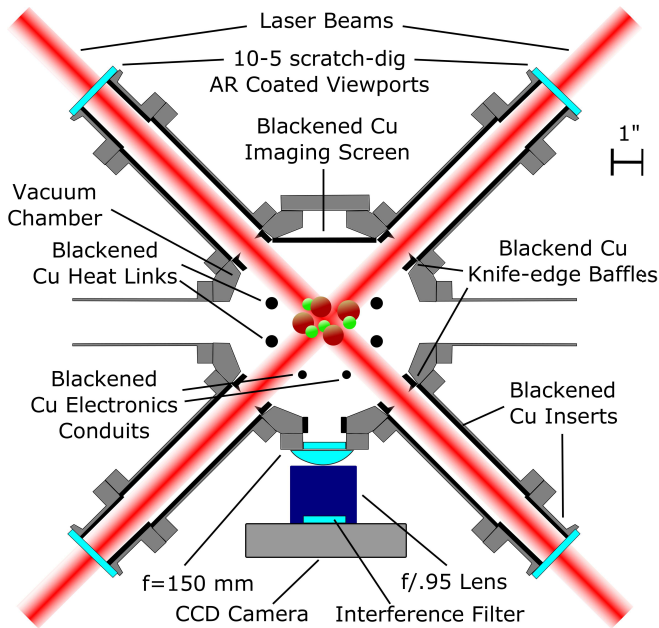}
        \caption{(Color online) Schematic of vacuum chamber used to detect SrF molecules in a magneto-optical trap using laser-induced fluorescence.  Several blackened Cu surfaces are used to reduce scattered light.  Blackened Cu heat links and electronics conduits are placed between high-power in-vacuum circuit boards.  The thickness of blackened surfaces are exaggerated for clarity.}\label{fig:motchamber}
\end{figure}

Fig.\,\ref{fig:examples}a shows stock and sandblasted copper samples before and after blackening for tests in this work.  Copper masked by adhesive PTFE tape is shown in Fig.\,\ref{fig:examples}e.  Recently, we have used blackened copper surfaces (Fig.\,\ref{fig:examples}b-d) to reduce background scattered light in experiments imaging magneto-optically trapped SrF molecules by LIF detection \cite{Barry2014,McCarron2015,Norrgard2015}.  A schematic of the detection region used in Ref.\,\cite{Norrgard2015} is shown in Fig.\,\ref{fig:motchamber}.   A 14\,mm 1/$e^2$ diameter laser beam (the beam is clipped by 23\,mm diameter optics external to the vacuum chamber)  containing 110\,mW of $\lambda$\,=\,663\,nm light is passed 6 times through the vacuum chamber (once each along three orthogonal axes and then retroreflected).  Excitation by the $\approx$\,660\,mW of $\lambda$\,=\,663\,nm light in the center of the chamber induces fluorescence from the trapped molecules at the same wavelength.  Light is collected by a 2''-diameter, 150-mm-focal-length spherical singlet lens placed directly outside the vacuum chamber, followed by a 50-mm-focal-length, f/0.95 camera lens.   The magnification factor at the imaging position is $M$\,=\,0.45.  Directly opposite the camera is a blackened Cu screen against which the molecules are imaged.  The total light detection efficiency for our system (including geometry, optical losses, and camera quantum efficiency) is 0.4\% \cite{Barry2014}.

 Scattering from the viewports is minimized by using high optical quality (10-5 scratch dig) glass with an anti-reflection V-coating at 663\,nm. Standard vacuum nipples with blackened copper sheet inserts (Fig.\,\ref{fig:examples}b) reduce stray light in two ways.  First, they reduce the power of light scattered directly from viewport surface imperfections into the LIF detection region, simply by placing the viewports further from the imaging chamber ($\approx$\,10'') than they would be in the absence of the nipple ($\approx$\,5'').  Second, the blackened inserts increase the likelihood that light incident on the walls of the nipple (either directly or through scattering) is absorbed before reaching the imaging region. Background light is additionally reduced by baffles \cite{Scoles1988} formed by blackened Cu rings with a knife-edged inner diameter machined to be 26\,mm, which sit inside the imaging chamber flush with the vacuum-sealing Cu gasket.

Fig.\,\ref{fig:examples}c shows a blackened copper disk for stray light absorption (not shown in Fig.\,\ref{fig:motchamber}) which covered the entire bottom of the molecule imaging chamber in \cite{Barry2014, McCarron2015}. A hole in the center provides optical access for the two vertical passes of the 663\,nm light.  For the work in Ref.\,\cite{Norrgard2015}, this disk was replaced with the assembly shown in Fig.\,\ref{fig:examples}d, where blackened copper rods serve as heat conduits (thick outer rods) for in-vacuum high-power circuit boards.  We found that the unmodified blackened surface does not allow for good thermal contact.  However, deliberately crushing the CuO dendrites on the thermally connected surface and applying a 0.004"-thick indium thermal interface layer (TIL) gave nearly the same thermal conductivity as obtained with a typical method for making heat links with unblackened copper (sanding the copper surface with 2000 grit sandpaper and applying an indium TIL).  Using an indium TIL without crushing the dendrites gave reduced conductivity (roughly 2$\times$ less), similar to that of sanded copper without indium present.

%I think this needs to be fleshed out a bit more.  I would say
%--give a figure with the optical setup around the MOT including nipples, locations of blackened surfaces, and optical collection setup.

 %Specify the laser power & beam diameter, and collection efficiency and/or f/# of collection optics.
%--give a number for how many scattered photons are seen in the camera (or in the region of interest, with specified size) and compare it to e.g. dark counts/read noise.

During normal trap operation, we measure the maximum photon scattering rate per molecule to be $R_{\rm{sc}}$\,=\,2.5$\times$\,10$^6$\,s$^{-1}$.   Analysis is restricted to a region of interest of typically $50 \times 50$ pixels ($3.2 \times 3.2$\,mm) where the majority of the LIF signal is imaged.  The background scattered light signal is typically $B = 1.5\times 10^4$ e$^-$/pixel/s.  For a typical $t_{\rm{e}}$\,=\,60\,ms exposure, the noise is usually limited by shot noise in this background, at the level $N_{\rm{B}} \approx$\,30\,e$^-$/pixel.  Under these conditions, the laser intensity noise, camera read noise, and camera dark current are approximately 4, 8 and 10$^4$ times smaller than $N_{\rm{B}}$, respectively. Integrating over the region of interest gives a signal-to-noise ratio (SNR) of 0.39 per molecule for a single trap loading shot.  With this system, we have observed MOTs of as few as 17 molecules \cite{Norrgard2015}, with SNR\,$\approx$\,110 when averaging over 300 successive shots.

%In some sense this is the take-home message of this paper, so we should emphasize it!  The statement that "we can detect as few at 17 SrF molecules" is good, but doesn't quite tell the story of just how much the scatter was suppressed here.

\section{Conclusion and Outlook}
We have used a procedure to produce blackened copper surfaces \cite{Meyer1944}, suitable for use in UHV applications.  We measure these surfaces to be only a few times more reflective than commercial UHV light-absorbing foil.  While such foils are available with low-outgassing adhesives, they typically have strict temperature requirements (40\,$^{\circ}$C\,$\lesssim$\,$T$\,$\lesssim$\,150\,$^{\circ}$C).  Blackened copper requires no adhesive, can be baked at high temperature in vacuum, and can be applied chemically to complex surfaces as well as to flexible foils and sheets.  Sandblasting prior to blackening is found to decrease diffuse back scatter, increase diffuse forward scatter, and remove specular reflection.

The advantages of vacuum-compatibility, rapid prototyping, and low cost make this technique ideally suited for scattered light suppression in experiments with LIF signals from a gas-phase source.  Recently, blackened copper surfaces installed in our vacuum chamber reduced stray light noise to a level sufficient to detect a small handful of molecules via LIF \cite{Norrgard2015}.  We also plan to use this technique to reduce scattered light background in various types of precision measurements using LIF from molecular beams \cite{Baron2014,Cahn2014,Hunter2012}, and we expect this technology will be useful in similar experiments.

% Blackening Method

%Reflectivity Measurements

%Thermal Conductivity

%Conclusions and Outlook

%The authors thank J.F. Barry, E.R. Edwards, D.J. McCarron, and M.H. Steinecker for contributions toward the construction of the vacuum chamber.
\begin{acknowledgments}
The authors thank E.R. Edwards for contributions toward the construction of the vacuum chamber. The authors acknowledge financial support from ARO and ARO (MURI). EBN acknowledges funding from the NSF GRFP. NS acknowledges funding from Yale College Dean's Research Fellowship.
\end{acknowledgments}
\bibliography{thebib}

\begin{thebibliography}{24}%
\makeatletter
\providecommand \@ifxundefined [1]{%
 \@ifx{#1\undefined}
}%
\providecommand \@ifnum [1]{%
 \ifnum #1\expandafter \@firstoftwo
 \else \expandafter \@secondoftwo
 \fi
}%
\providecommand \@ifx [1]{%
 \ifx #1\expandafter \@firstoftwo
 \else \expandafter \@secondoftwo
 \fi
}%
\providecommand \natexlab [1]{#1}%
\providecommand \enquote  [1]{``#1''}%
\providecommand \bibnamefont  [1]{#1}%
\providecommand \bibfnamefont [1]{#1}%
\providecommand \citenamefont [1]{#1}%
\providecommand \href@noop [0]{\@secondoftwo}%
\providecommand \href [0]{\begingroup \@sanitize@url \@href}%
\providecommand \@href[1]{\@@startlink{#1}\@@href}%
\providecommand \@@href[1]{\endgroup#1\@@endlink}%
\providecommand \@sanitize@url [0]{\catcode `\\12\catcode `\$12\catcode
  `\&12\catcode `\#12\catcode `\^12\catcode `\_12\catcode `\%12\relax}%
\providecommand \@@startlink[1]{}%
\providecommand \@@endlink[0]{}%
\providecommand \url  [0]{\begingroup\@sanitize@url \@url }%
\providecommand \@url [1]{\endgroup\@href {#1}{\urlprefix }}%
\providecommand \urlprefix  [0]{URL }%
\providecommand \Eprint [0]{\href }%
\providecommand \doibase [0]{http://dx.doi.org/}%
\providecommand \selectlanguage [0]{\@gobble}%
\providecommand \bibinfo  [0]{\@secondoftwo}%
\providecommand \bibfield  [0]{\@secondoftwo}%
\providecommand \translation [1]{[#1]}%
\providecommand \BibitemOpen [0]{}%
\providecommand \bibitemStop [0]{}%
\providecommand \bibitemNoStop [0]{.\EOS\space}%
\providecommand \EOS [0]{\spacefactor3000\relax}%
\providecommand \BibitemShut  [1]{\csname bibitem#1\endcsname}%
\let\auto@bib@innerbib\@empty
%</preamble>
\bibitem [{\citenamefont {Kinsey}(1977)}]{Kinsey1977}%
  \BibitemOpen
  \bibfield  {author} {\bibinfo {author} {\bibfnamefont {J.~L.}\ \bibnamefont
  {Kinsey}},\ }\href {\doibase 10.1146/annurev.pc.28.100177.002025} {\bibfield
  {journal} {\bibinfo  {journal} {Annual Review of Physical Chemistry}\
  }\textbf {\bibinfo {volume} {28}},\ \bibinfo {pages} {349} (\bibinfo {year}
  {1977})}\BibitemShut {NoStop}%
\bibitem [{\citenamefont {Scoles}(1988)}]{Scoles1988}%
  \BibitemOpen
  \bibfield  {author} {\bibinfo {author} {\bibfnamefont {G.}~\bibnamefont
  {Scoles}},\ }\href@noop {} {\emph {\bibinfo {title} {Atomic and Molecular
  Beam Methods}}}\ (\bibinfo  {publisher} {Oxford University Press},\ \bibinfo
  {address} {New York},\ \bibinfo {year} {1988})\BibitemShut {NoStop}%
\bibitem [{\citenamefont {Pauly}(2000)}]{Pauly2000}%
  \BibitemOpen
  \bibfield  {author} {\bibinfo {author} {\bibfnamefont {H.}~\bibnamefont
  {Pauly}},\ }\href {http://dx.doi.org/10.1002/0470027320.s2405} {\emph
  {\bibinfo {title} {Atom, Molecule, and Cluster Beams I: Basic Theory,
  Production, and Detection of Thermal Energy Beams}}}\ (\bibinfo  {publisher}
  {Springer-Verlag},\ \bibinfo {year} {2000})\BibitemShut {NoStop}%
\bibitem [{\citenamefont {Demtr\"{o}der}(2015)}]{Demtroder2015}%
  \BibitemOpen
  \bibfield  {author} {\bibinfo {author} {\bibfnamefont {W.}~\bibnamefont
  {Demtr\"{o}der}},\ }\href {http://dx.doi.org/10.1007/978-3-662-44641-6}
  {\emph {\bibinfo {title} {Laser Spectroscopy 2: Experimental Techniques}}},\
  \bibinfo {edition} {5th}\ ed.\ (\bibinfo  {publisher} {Springer-Verlag},\
  \bibinfo {year} {2015})\BibitemShut {NoStop}%
\bibitem [{\citenamefont {Gross}\ and\ \citenamefont
  {Bloch}(2015)}]{Gross2015}%
  \BibitemOpen
  \bibfield  {author} {\bibinfo {author} {\bibfnamefont {C.}~\bibnamefont
  {Gross}}\ and\ \bibinfo {author} {\bibfnamefont {I.}~\bibnamefont {Bloch}},\
  }\enquote {\bibinfo {title} {Microscopy of many-body states in optical
  lattices},}\ in\ \href@noop {} {\emph {\bibinfo {booktitle} {Annual Review of
  Cold Atoms and Molecules}}},\ Vol.~\bibinfo {volume} {3}\ (\bibinfo
  {publisher} {World Scientific Publishing Co.},\ \bibinfo {year}
  {2015})\BibitemShut {NoStop}%
\bibitem [{\citenamefont {Moerner}\ and\ \citenamefont
  {Fromm}(2003)}]{Moerner2003}%
  \BibitemOpen
  \bibfield  {author} {\bibinfo {author} {\bibfnamefont {W.~E.}\ \bibnamefont
  {Moerner}}\ and\ \bibinfo {author} {\bibfnamefont {D.~P.}\ \bibnamefont
  {Fromm}},\ }\href {\doibase http://dx.doi.org/10.1063/1.1589587} {\bibfield
  {journal} {\bibinfo  {journal} {Review of Scientific Instruments}\ }\textbf
  {\bibinfo {volume} {74}},\ \bibinfo {pages} {3597} (\bibinfo {year}
  {2003})}\BibitemShut {NoStop}%
\bibitem [{\citenamefont {Leibfried}\ \emph {et~al.}(2003)\citenamefont
  {Leibfried}, \citenamefont {Blatt}, \citenamefont {Monroe},\ and\
  \citenamefont {Wineland}}]{Leibfried2003}%
  \BibitemOpen
  \bibfield  {author} {\bibinfo {author} {\bibfnamefont {D.}~\bibnamefont
  {Leibfried}}, \bibinfo {author} {\bibfnamefont {R.}~\bibnamefont {Blatt}},
  \bibinfo {author} {\bibfnamefont {C.}~\bibnamefont {Monroe}}, \ and\ \bibinfo
  {author} {\bibfnamefont {D.}~\bibnamefont {Wineland}},\ }\href {\doibase
  10.1103/RevModPhys.75.281} {\bibfield  {journal} {\bibinfo  {journal} {Rev.
  Mod. Phys.}\ }\textbf {\bibinfo {volume} {75}},\ \bibinfo {pages} {281}
  (\bibinfo {year} {2003})}\BibitemShut {NoStop}%
\bibitem [{\citenamefont {Hall}\ \emph {et~al.}(1984)\citenamefont {Hall},
  \citenamefont {Liu}, \citenamefont {McAuliffe}, \citenamefont {Giese},\ and\
  \citenamefont {Gentry}}]{Hall1984}%
  \BibitemOpen
  \bibfield  {author} {\bibinfo {author} {\bibfnamefont {G.}~\bibnamefont
  {Hall}}, \bibinfo {author} {\bibfnamefont {K.}~\bibnamefont {Liu}}, \bibinfo
  {author} {\bibfnamefont {M.~J.}\ \bibnamefont {McAuliffe}}, \bibinfo {author}
  {\bibfnamefont {C.~F.}\ \bibnamefont {Giese}}, \ and\ \bibinfo {author}
  {\bibfnamefont {W.~R.}\ \bibnamefont {Gentry}},\ }\href {\doibase
  http://dx.doi.org/10.1063/1.447660} {\bibfield  {journal} {\bibinfo
  {journal} {The Journal of Chemical Physics}\ }\textbf {\bibinfo {volume}
  {81}},\ \bibinfo {pages} {5577} (\bibinfo {year} {1984})}\BibitemShut
  {NoStop}%
\bibitem [{\citenamefont {{Barry}}\ \emph {et~al.}(2012)\citenamefont
  {{Barry}}, \citenamefont {{Shuman}}, \citenamefont {{Norrgard}},\ and\
  \citenamefont {{DeMille}}}]{Barry2012}%
  \BibitemOpen
  \bibfield  {author} {\bibinfo {author} {\bibfnamefont {J.~F.}\ \bibnamefont
  {{Barry}}}, \bibinfo {author} {\bibfnamefont {E.~S.}\ \bibnamefont
  {{Shuman}}}, \bibinfo {author} {\bibfnamefont {E.~B.}\ \bibnamefont
  {{Norrgard}}}, \ and\ \bibinfo {author} {\bibfnamefont {D.}~\bibnamefont
  {{DeMille}}},\ }\href {\doibase 10.1103/PhysRevLett.108.103002} {\bibfield
  {journal} {\bibinfo  {journal} {Phys. Rev. Lett.}\ }\textbf {\bibinfo
  {volume} {108}},\ \bibinfo {eid} {103002} (\bibinfo {year}
  {2012})}\BibitemShut {NoStop}%
\bibitem [{\citenamefont {{Shuman}}\ \emph {et~al.}(2010)\citenamefont
  {{Shuman}}, \citenamefont {{Barry}},\ and\ \citenamefont
  {{DeMille}}}]{Shuman2010}%
  \BibitemOpen
  \bibfield  {author} {\bibinfo {author} {\bibfnamefont {E.~S.}\ \bibnamefont
  {{Shuman}}}, \bibinfo {author} {\bibfnamefont {J.~F.}\ \bibnamefont
  {{Barry}}}, \ and\ \bibinfo {author} {\bibfnamefont {D.}~\bibnamefont
  {{DeMille}}},\ }\href {\doibase 10.1038/nature09443} {\bibfield  {journal}
  {\bibinfo  {journal} {Nature}\ }\textbf {\bibinfo {volume} {467}},\ \bibinfo
  {pages} {820} (\bibinfo {year} {2010})}\BibitemShut {NoStop}%
\bibitem [{\citenamefont {Barry}\ \emph {et~al.}(2014)\citenamefont {Barry},
  \citenamefont {McCarron}, \citenamefont {Norrgard}, \citenamefont
  {Steinecker},\ and\ \citenamefont {DeMille}}]{Barry2014}%
  \BibitemOpen
  \bibfield  {author} {\bibinfo {author} {\bibfnamefont {J.~F.}\ \bibnamefont
  {Barry}}, \bibinfo {author} {\bibfnamefont {D.~J.}\ \bibnamefont {McCarron}},
  \bibinfo {author} {\bibfnamefont {E.~N.}\ \bibnamefont {Norrgard}}, \bibinfo
  {author} {\bibfnamefont {M.~H.}\ \bibnamefont {Steinecker}}, \ and\ \bibinfo
  {author} {\bibfnamefont {D.}~\bibnamefont {DeMille}},\ }\href {\doibase
  10.1038/nature13634} {\bibfield  {journal} {\bibinfo  {journal} {Nature}\
  }\textbf {\bibinfo {volume} {512}},\ \bibinfo {pages} {286} (\bibinfo {year}
  {2014})}\BibitemShut {NoStop}%
\bibitem [{\citenamefont {McCarron}\ \emph {et~al.}(2015)\citenamefont
  {McCarron}, \citenamefont {Norrgard}, \citenamefont {Steinecker},\ and\
  \citenamefont {DeMille}}]{McCarron2015}%
  \BibitemOpen
  \bibfield  {author} {\bibinfo {author} {\bibfnamefont {D.~J.}\ \bibnamefont
  {McCarron}}, \bibinfo {author} {\bibfnamefont {E.~B.}\ \bibnamefont
  {Norrgard}}, \bibinfo {author} {\bibfnamefont {M.~H.}\ \bibnamefont
  {Steinecker}}, \ and\ \bibinfo {author} {\bibfnamefont {D.}~\bibnamefont
  {DeMille}},\ }\href {http://stacks.iop.org/1367-2630/17/i=3/a=035014}
  {\bibfield  {journal} {\bibinfo  {journal} {New Journal of Physics}\ }\textbf
  {\bibinfo {volume} {17}},\ \bibinfo {pages} {035014} (\bibinfo {year}
  {2015})}\BibitemShut {NoStop}%
\bibitem [{\citenamefont {{Norrgard}}\ \emph {et~al.}(2015)\citenamefont
  {{Norrgard}}, \citenamefont {{McCarron}}, \citenamefont {{Steinecker}},
  \citenamefont {{Tarbutt}},\ and\ \citenamefont {{DeMille}}}]{Norrgard2015}%
  \BibitemOpen
  \bibfield  {author} {\bibinfo {author} {\bibfnamefont {E.~B.}\ \bibnamefont
  {{Norrgard}}}, \bibinfo {author} {\bibfnamefont {D.~J.}\ \bibnamefont
  {{McCarron}}}, \bibinfo {author} {\bibfnamefont {M.~H.}\ \bibnamefont
  {{Steinecker}}}, \bibinfo {author} {\bibfnamefont {M.~R.}\ \bibnamefont
  {{Tarbutt}}}, \ and\ \bibinfo {author} {\bibfnamefont {D.}~\bibnamefont
  {{DeMille}}},\ }\href@noop {} {\  (\bibinfo {year} {2015})},\ \Eprint
  {http://arxiv.org/abs/1511.00930} {arXiv:1511.00930} \BibitemShut {NoStop}%
\bibitem [{\citenamefont {Kirsch}\ and\ \citenamefont
  {Ekerdt}(2001)}]{Kirsch2001}%
  \BibitemOpen
  \bibfield  {author} {\bibinfo {author} {\bibfnamefont {P.~D.}\ \bibnamefont
  {Kirsch}}\ and\ \bibinfo {author} {\bibfnamefont {J.~G.}\ \bibnamefont
  {Ekerdt}},\ }\href {\doibase http://dx.doi.org/10.1063/1.1403675} {\bibfield
  {journal} {\bibinfo  {journal} {Journal of Applied Physics}\ }\textbf
  {\bibinfo {volume} {90}},\ \bibinfo {pages} {4256} (\bibinfo {year}
  {2001})}\BibitemShut {NoStop}%
\bibitem [{\citenamefont {Meyer}(1944)}]{Meyer1944}%
  \BibitemOpen
  \bibfield  {author} {\bibinfo {author} {\bibfnamefont {W.}~\bibnamefont
  {Meyer}},\ }\href {http://www.google.com/patents/US2364993} {\enquote
  {\bibinfo {title} {Process for blackening copper or copper alloy surfaces},}\
  } (\bibinfo {year} {1944}),\ \bibinfo {note} {{US} Patent
  2,364,993}\BibitemShut {NoStop}%
\bibitem [{\citenamefont {Filipi\v{c}}\ and\ \citenamefont
  {Cvelbar}(2012)}]{Filipic2012}%
  \BibitemOpen
  \bibfield  {author} {\bibinfo {author} {\bibfnamefont {G.}~\bibnamefont
  {Filipi\v{c}}}\ and\ \bibinfo {author} {\bibfnamefont {U.}~\bibnamefont
  {Cvelbar}},\ }\href {http://stacks.iop.org/0957-4484/23/i=19/a=194001}
  {\bibfield  {journal} {\bibinfo  {journal} {Nanotechnology}\ }\textbf
  {\bibinfo {volume} {23}},\ \bibinfo {pages} {194001} (\bibinfo {year}
  {2012})}\BibitemShut {NoStop}%
\bibitem [{\citenamefont {{Sandvik Materials Technology}}()}]{Sandvik}%
  \BibitemOpen
  \bibfield  {author} {\bibinfo {author} {\bibnamefont {{Sandvik Materials
  Technology}}},\ }\href
  {http://smt.sandvik.com/en/materials-center/corrosion-tables/sodium-hydroxide/}
  {\enquote {\bibinfo {title} {Sodium hydroxide corrosion tables},}\ }\bibinfo
  {note} {Available online at
  http://smt.sandvik.com/en/materials-center/corrosion-tables/sodium-hydroxide/
  (viewed 12 Jan 2016)}\BibitemShut {NoStop}%
\bibitem [{\citenamefont {Lebbai}\ \emph {et~al.}(2003)\citenamefont {Lebbai},
  \citenamefont {Kim}, \citenamefont {Szeto}, \citenamefont {Yuen},\ and\
  \citenamefont {Tong}}]{Lebbai2003}%
  \BibitemOpen
  \bibfield  {author} {\bibinfo {author} {\bibfnamefont {M.}~\bibnamefont
  {Lebbai}}, \bibinfo {author} {\bibfnamefont {J.-K.}\ \bibnamefont {Kim}},
  \bibinfo {author} {\bibfnamefont {W.}~\bibnamefont {Szeto}}, \bibinfo
  {author} {\bibfnamefont {M.}~\bibnamefont {Yuen}}, \ and\ \bibinfo {author}
  {\bibfnamefont {P.}~\bibnamefont {Tong}},\ }\href {\doibase
  10.1007/s11664-003-0142-y} {\bibfield  {journal} {\bibinfo  {journal}
  {Journal of Electronic Materials}\ }\textbf {\bibinfo {volume} {32}},\
  \bibinfo {pages} {558} (\bibinfo {year} {2003})}\BibitemShut {NoStop}%
\bibitem [{\citenamefont {Shao}\ and\ \citenamefont
  {Zangari}(2009)}]{Shao2009}%
  \BibitemOpen
  \bibfield  {author} {\bibinfo {author} {\bibfnamefont {W.}~\bibnamefont
  {Shao}}\ and\ \bibinfo {author} {\bibfnamefont {G.}~\bibnamefont {Zangari}},\
  }\href {http://dx.doi.org/10.1021/jp8095456} {\bibfield  {journal} {\bibinfo
  {journal} {The Journal of Physical Chemistry C}\ }\textbf {\bibinfo {volume}
  {113}},\ \bibinfo {pages} {10097} (\bibinfo {year} {2009})}\BibitemShut
  {NoStop}%
\bibitem [{\citenamefont {Hu}\ \emph {et~al.}(2010)\citenamefont {Hu},
  \citenamefont {Huang}, \citenamefont {Want}, \citenamefont {Liu},
  \citenamefont {Jiang}, \citenamefont {Ding}, \citenamefont {Ji},\ and\
  \citenamefont {Li}}]{Hu2010}%
  \BibitemOpen
  \bibfield  {author} {\bibinfo {author} {\bibfnamefont {Y.}~\bibnamefont
  {Hu}}, \bibinfo {author} {\bibfnamefont {X.}~\bibnamefont {Huang}}, \bibinfo
  {author} {\bibfnamefont {K.}~\bibnamefont {Want}}, \bibinfo {author}
  {\bibfnamefont {J.}~\bibnamefont {Liu}}, \bibinfo {author} {\bibfnamefont
  {J.}~\bibnamefont {Jiang}}, \bibinfo {author} {\bibfnamefont
  {R.}~\bibnamefont {Ding}}, \bibinfo {author} {\bibfnamefont {X.}~\bibnamefont
  {Ji}}, \ and\ \bibinfo {author} {\bibfnamefont {X.}~\bibnamefont {Li}},\
  }\href {\doibase http://dx.doi.org/10.1016/j.jssc.2010.01.013} {\bibfield
  {journal} {\bibinfo  {journal} {Journal of Solid State Chemistry}\ }\textbf
  {\bibinfo {volume} {183}},\ \bibinfo {pages} {662 } (\bibinfo {year}
  {2010})}\BibitemShut {NoStop}%
\bibitem [{\citenamefont {Xia}\ \emph {et~al.}(2014)\citenamefont {Xia},
  \citenamefont {Pu}, \citenamefont {Liu}, \citenamefont {Liang}, \citenamefont
  {Liu}, \citenamefont {Li},\ and\ \citenamefont {Yu}}]{Xia2014}%
  \BibitemOpen
  \bibfield  {author} {\bibinfo {author} {\bibfnamefont {Y.}~\bibnamefont
  {Xia}}, \bibinfo {author} {\bibfnamefont {X.}~\bibnamefont {Pu}}, \bibinfo
  {author} {\bibfnamefont {J.}~\bibnamefont {Liu}}, \bibinfo {author}
  {\bibfnamefont {J.}~\bibnamefont {Liang}}, \bibinfo {author} {\bibfnamefont
  {P.}~\bibnamefont {Liu}}, \bibinfo {author} {\bibfnamefont {X.}~\bibnamefont
  {Li}}, \ and\ \bibinfo {author} {\bibfnamefont {X.}~\bibnamefont {Yu}},\
  }\href {\doibase 10.1039/C4TA00097H} {\bibfield  {journal} {\bibinfo
  {journal} {J. Mater. Chem. A}\ }\textbf {\bibinfo {volume} {2}},\ \bibinfo
  {pages} {6796} (\bibinfo {year} {2014})}\BibitemShut {NoStop}%
\bibitem [{\citenamefont {{The ACME Collaboration, Baron, J.}}\ \emph
  {et~al.}(2014)\citenamefont {{The ACME Collaboration, Baron, J.}},
  \citenamefont {Baron}, \citenamefont {Campbell}, \citenamefont {DeMille},
  \citenamefont {Doyle}, \citenamefont {Gabrielse}, \citenamefont {Gurevich},
  \citenamefont {Hess}, \citenamefont {Hutzler}, \citenamefont {Kirilov},
  \citenamefont {Kozyryev}, \citenamefont {O'Leary}, \citenamefont {Panda},
  \citenamefont {Parsons}, \citenamefont {Petrik}, \citenamefont {Spaun},
  \citenamefont {Vutha},\ and\ \citenamefont {West}}]{Baron2014}%
  \BibitemOpen
  \bibfield  {author} {\bibinfo {author} {\bibnamefont {{The ACME
  Collaboration, Baron, J.}}}, \bibinfo {author} {\bibfnamefont
  {J.}~\bibnamefont {Baron}}, \bibinfo {author} {\bibfnamefont {W.~C.}\
  \bibnamefont {Campbell}}, \bibinfo {author} {\bibfnamefont {D.}~\bibnamefont
  {DeMille}}, \bibinfo {author} {\bibfnamefont {J.~M.}\ \bibnamefont {Doyle}},
  \bibinfo {author} {\bibfnamefont {G.}~\bibnamefont {Gabrielse}}, \bibinfo
  {author} {\bibfnamefont {Y.~V.}\ \bibnamefont {Gurevich}}, \bibinfo {author}
  {\bibfnamefont {P.~W.}\ \bibnamefont {Hess}}, \bibinfo {author}
  {\bibfnamefont {N.~R.}\ \bibnamefont {Hutzler}}, \bibinfo {author}
  {\bibfnamefont {E.}~\bibnamefont {Kirilov}}, \bibinfo {author} {\bibfnamefont
  {I.}~\bibnamefont {Kozyryev}}, \bibinfo {author} {\bibfnamefont {B.~R.}\
  \bibnamefont {O'Leary}}, \bibinfo {author} {\bibfnamefont {C.~D.}\
  \bibnamefont {Panda}}, \bibinfo {author} {\bibfnamefont {M.~F.}\ \bibnamefont
  {Parsons}}, \bibinfo {author} {\bibfnamefont {E.~S.}\ \bibnamefont {Petrik}},
  \bibinfo {author} {\bibfnamefont {B.}~\bibnamefont {Spaun}}, \bibinfo
  {author} {\bibfnamefont {A.~C.}\ \bibnamefont {Vutha}}, \ and\ \bibinfo
  {author} {\bibfnamefont {A.~D.}\ \bibnamefont {West}},\ }\href {\doibase
  10.1126/science.1248213} {\ \textbf {\bibinfo {volume} {343}},\ \bibinfo
  {pages} {269} (\bibinfo {year} {2014})}\BibitemShut {NoStop}%
\bibitem [{\citenamefont {Cahn}\ \emph {et~al.}(2014)\citenamefont {Cahn},
  \citenamefont {Ammon}, \citenamefont {Kirilov}, \citenamefont {Gurevich},
  \citenamefont {Murphree}, \citenamefont {Paolino}, \citenamefont {Rahmlow},
  \citenamefont {Kozlov},\ and\ \citenamefont {DeMille}}]{Cahn2014}%
  \BibitemOpen
  \bibfield  {author} {\bibinfo {author} {\bibfnamefont {S.~B.}\ \bibnamefont
  {Cahn}}, \bibinfo {author} {\bibfnamefont {J.}~\bibnamefont {Ammon}},
  \bibinfo {author} {\bibfnamefont {E.}~\bibnamefont {Kirilov}}, \bibinfo
  {author} {\bibfnamefont {Y.~V.}\ \bibnamefont {Gurevich}}, \bibinfo {author}
  {\bibfnamefont {D.}~\bibnamefont {Murphree}}, \bibinfo {author}
  {\bibfnamefont {R.}~\bibnamefont {Paolino}}, \bibinfo {author} {\bibfnamefont
  {D.~A.}\ \bibnamefont {Rahmlow}}, \bibinfo {author} {\bibfnamefont {M.~G.}\
  \bibnamefont {Kozlov}}, \ and\ \bibinfo {author} {\bibfnamefont
  {D.}~\bibnamefont {DeMille}},\ }\href {\doibase
  10.1103/PhysRevLett.112.163002} {\bibfield  {journal} {\bibinfo  {journal}
  {Phys. Rev. Lett.}\ }\textbf {\bibinfo {volume} {112}},\ \bibinfo {pages}
  {163002} (\bibinfo {year} {2014})}\BibitemShut {NoStop}%
\bibitem [{\citenamefont {{Hunter}}\ \emph {et~al.}(2012)\citenamefont
  {{Hunter}}, \citenamefont {{Peck}}, \citenamefont {{Greenspon}},
  \citenamefont {{Alam}},\ and\ \citenamefont {{DeMille}}}]{Hunter2012}%
  \BibitemOpen
  \bibfield  {author} {\bibinfo {author} {\bibfnamefont {L.~R.}\ \bibnamefont
  {{Hunter}}}, \bibinfo {author} {\bibfnamefont {S.~K.}\ \bibnamefont
  {{Peck}}}, \bibinfo {author} {\bibfnamefont {A.~S.}\ \bibnamefont
  {{Greenspon}}}, \bibinfo {author} {\bibfnamefont {S.~S.}\ \bibnamefont
  {{Alam}}}, \ and\ \bibinfo {author} {\bibfnamefont {D.}~\bibnamefont
  {{DeMille}}},\ }\href {\doibase 10.1103/PhysRevA.85.012511} {\bibfield
  {journal} {\bibinfo  {journal} {\pra}\ }\textbf {\bibinfo {volume} {85}},\
  \bibinfo {eid} {012511} (\bibinfo {year} {2012})}\BibitemShut {NoStop}%
\end{thebibliography}%

\clearpage

\end{document}